# Giant Piezoelectricity in Monolayer Group IV Monochalcogenides: SnSe, SnS, GeSe and GeS

Ruixiang Fei,[1] Wenbin Li,[2] Ju Li,[3] and Li Yang[1*]

[1] Department of Physics, Washington University, St Louis, MO 63130, United States

[2] Research Laboratory of Electronics, Massachusetts Institute of Technology, Cambridge, MA 02139, United States

[3] Department of Nuclear Science and Engineering and Department of Materials Science and Engineering, Massachusetts Institute of Technology, Cambridge, MA 02139, United States

**ABSTRACT:** We predict enormous piezoelectric effects in intrinsic monolayer group IV monochalcogenides (MX, M=Sn or Ge, X=Se or S), including SnSe, SnS, GeSe and GeS. Using first-principle simulations based on the modern theory of polarization, we find that their piezoelectric coefficients are about one to two orders of magnitude larger than those of other 2D materials, such as $MoS_2$ and GaSe, and bulk quartz and AlN which are widely used in industry. This enhancement is a result of the unique "puckered" $C_{2v}$ symmetry and weaker chemical bonds of monolayer group IV monochalcogenides. Given the achieved experimental advances in fabrication of monolayers, their flexible character and ability to withstand enormous strain, these 2D structures with giant piezoelectric effects may be promising for a broad range of applications, such as nano-sized sensors, piezotronics, and energy harvesting in portable electronic devices.

*Introduction:* Piezoelectric materials, which convert mechanical energy to electrical energy, have the advantages of large power densities and ease of application in sensors and energy harvesting,[1,2] For example, a widely used piezoelectric material is lead zirconate titanate $Pb[Zr_xTi_{1-x}]O_3$, a piezoceramic known as PZT. [3-5] However, the piezoceramic's brittle nature causes limitations in the sustainable strain.[6] Meanwhile, non-centrosymmetric wurtzite-structured semiconductors, such as ZnO, GaN and InN, are wildly used in piezotronic and piezo-phototronic devices [7-9]. In particular, their nanowires or nanobelts [10-12] are expected to be useful for electromechanical coupled sensors, nanoscale energy conversion for self-powered nano-devices, and harvesting energy from the environment.[10-13] However, compared to piezoceramics, the much smaller piezoelectric coefficients of wurtzite semiconductors limit the mechanical-electrical energy conversion efficiency.[7,8]

Recently two-dimensional (2D) materials have sparked interest for piezoelectric applications because of their high crystallinity and ability to withstand enormous strain. For those hexagonal structures with a $D_{6h}$ point group, such as boron nitride (*h*-BN) and many transition-metal dichalcogenides (TMDCs), as well as layered orthorhombic structure with a $D_{4h}$ point group, such as Group-III monochalcogenides, their symmetry is reduced to the $D_{3h}$ group when thinned down to monolayer. This breaks the inversion symmetry, as shown in Figs. 1 (a) and (b), and gives rise to piezoelectricity. Thus they were theoretically predicted to be intrinsically piezoelectric[14-16] and this idea has been demonstrated by experiments on $MoS_2$ monolayer [17-19]. Unfortunately, the piezoelectric effect is rather small, e.g., the measured piezoelectric coefficient $e_{11}$ of monolayer $MoS_2$ is only around $2.9x10^{-10}$ C/m [18], and the corresponding mechanical-electrical energy conversion rate is limited to be about 5%[17].

Therefore, finding flexible, stable, and efficient 2D piezoelectric materials is crucial. This motivates us to study another family of 2D semiconductors, the group IV monochalcogenides (MX, M=Sn or Ge, M=Se or S), i.e., SnSe, SnS, GeSe and GeS. Their atomic structure is presented in Fig. 1 (c) and (d), which exhibit a different symmetry, the $C_{2v}$ point group. We expect enhanced piezoelectricity due to the following reasons: 1) As showed in Fig. 1 (c), their stable monolayer structures are non-centrosymmetric, which allows them to be piezoelectric. 2) Their puckered $C_{2v}$ symmetry are much more flexible (softer) along the armchair direction, compared with other $D_{3h}$ symmetry materials. This can further enhance the piezoelectric effect because the structure is more sensitive to the applied stress. 3) Significant advances in fabrication techniques have been achieved, making our prediction meaningful for immediate applications. For example, monolayer and few-layer structures have been fabricated recently. [20]

In this Letter, we employ first-principles density functional theory (DFT) simulations to calculate piezoelectric effects of monolayer group-IV monochalcogenides. Compared with other 2D materials, the calculated elastic stiffness is substantially smaller and the polarization induced by stress is significantly larger. As a result, the piezoelectric effect of these monolayer materials is dramatically enhanced and the piezoelectric coefficient $d_{11}$ is about one to two orders of magnitude larger than that of 2D $MoS_2$, GaS, and bulk quartz and AlN, which have been widely used in the industry.[7,14,16] These intrinsic, giant piezoelectric materials represent a new class of nanomaterials that will allow for the next generation of ultra-sensitive mechanical detectors, energy conversion devices and consumer-touch sensors.

*Computational approaches:* The DFT calculations with the Perdew-Burke-Ernzerh (PBE) functional [21] have been carried out by using the Vienna Ab initio Simulation Package (VASP) with a plane wave basis set [22,23]and the projector-augmented wave method[24]. The plane-wave cutoff energy used is 600 eV. To facilitate calculations of unit-cell polarization under strain, we use an orthorhombic unit cell containing two M atoms and two X atoms, as indicated in Fig. 1 (c). The interlayer distance is set to be 20 Å to mimic suspended monolayers. The Brillouin zone integration is obtained by a 14x14x1 k-point grid. The convergence criteria for electronic and ionic relaxations are $10^{-6}$ eV and $10^{-3}$ eV/ Å, respectively. We use the "Berry-phase" theory of polarization to directly compute the electric polarization.[25-27] The change of polarization ($\Delta P$) occurs upon making an adiabatic change in the Kohn-Sham Hamiltonian of the crystal.

*Atomic structure:* The DFT-optimized monolayer and bulk structure parameters, i.e., the in-plane lattice constants *a* and *b*, are listed in table I. The corresponding experimental or previous DFT results of the bulk phase are listed as well.[28-36] We observe a similar trend as that in Ref. 37, in which the lattice constant *a* increases and the constant *b* decreases with increasing the number of layers for most group IV monochalcogenides, except for GeS. Additionally, our DFT calculations of monolayer structures are in good agreement with previous studies [37] These monolayers shall be stable. This is evidenced by recent, successful experimental fabrications [20] and theoretical phonon calculations.[46]

We have calculated the band structure of the group IV monochalcogenides, which is presented in supplementary information. All these materials exhibit an indirect band gap at the DFT level, which is also present in their bulk phases. We list the values of indirect band gaps and direct gaps in Table I. These DFT gap values are for reference purposes only, as excited-state calculations are needed for get the reliable band gap of MXs. According to our experience,[38,39] the quasiparticle band gaps of monolayer MX is usually 150% ~200% larger than the DFT values. Thus we expect that the band gaps of MX range from 1.2 eV to 2.7 eV, which are in a very useful range for electronic applications. Moreover, due to the confinement effects in monolayer structures, huge electron-hole interactions and excitonic effects are expected, which can substantially lower the optical absorption edge by a few hundred meV, making these materials promising for solar energy applications. [20,40]

Piezoelectric properties are ground-state properties associated with polarization. Thus DFT calculations are a suitable tool shown to reliably predict values. For example, the DFT calculated piezoelectric coefficients are in excellent agreement with experimental values for bulk GaN[45] and nanostructure. Very recently, experiments measured the piezoelectric coefficient $e_{11}=2.9\text{x}10^{-10}$ C/m for monolayer $MoS_2$, which is close to the DFT results ($3.6\times10^{-10}$ C/m) .[18] Therefore, we employ the same theoretical approach in this work.

*Elastic stiffness:* We first obtained the planar elastic stiffness coefficients $C_{11}$, $C_{22}$ and $C_{12}$ of MX monolayer by fitting the DFT-calculated unit-cell energy $U$ to a series of 2D strain states ($\varepsilon_{11}$,$\varepsilon_{22}$), based on the formula

$$C_{11}=\frac{1}{A_0}\frac{\partial^2 U}{\partial\varepsilon_{11}^2}\ ,C_{22}=\frac{1}{A_0}\frac{\partial^2 U}{\partial\varepsilon_{22}^2}\ ,C_{12}=\frac{1}{A_0}\frac{\partial^2 U}{\partial\varepsilon_{11}\partial\varepsilon_{22}} \qquad (1)$$

where $A_0$ is the unit-cell area at the zero strain. Due to the existence of mirror symmetry along the zigzag direction (y direction) in MX structures, at the small strain limit, we can write

$$\Delta u(\varepsilon_{11},\varepsilon_{22})=\frac{1}{2}C_{11}\varepsilon_{11}^2+\frac{1}{2}C_{22}\varepsilon_{22}^2+C_{12}\varepsilon_{11}\varepsilon_{22} \qquad (2)$$

where $\Delta u(\varepsilon_{11}, \varepsilon_{22}) = [U(\varepsilon_{11}, \varepsilon_{22}) - U(\varepsilon_{11} = 0, \varepsilon_{22} = 0)]/A_0$ is the change of unit-cell energy per area. We carry out the strain energy calculation on an 11×11 grid with $\varepsilon_{11}$ and $\varepsilon_{22}$ ranging from -0.005 to 0.005 in steps of 0.001. The atomic positions in the strained unit cell are allowed to be fully relaxed. Following definitions of previous works,[18] the coefficients, $C_{11}$, $C_{22}$, and $C_{12}$, which are calculated using a fully relaxed final atomic configuration, are called relaxed-ion stiffness coefficients, which are experimentally relevant. In contrast, if the atomic positions are held fixed when applying unit-cell strain, the so-called clamped-ion coefficients, which are of theoretical interest, can be calculated as well.

Table II summarizes the clamped and relaxed-ion stiffness coefficients for the four types of $C_{2v}$ symmetry MX monolayers. Additionally, we have also listed the elastic stiffness of another two typical $D_{3h}$ symmetry piezoelectric materials, $MoS_2$ [14] and GaSe[16], which belong to the TMDC and group-III monochalcogenide classes, respectively. According to the structures shown in Fig. 1 (c) and (d), the group IV monochalcogenide are soft along the armchair (x) direction. This is consistent with our DFT results in Table II. In particular, for both clamped and relax-ion cases, the elastic stiff nesses ($C_{11}$) of group-IV monochalcogenide is about 4~6 times smaller than that of $MoS_2$ and GaSe. This will significantly enhance the piezoelectric effects. An unexpected result from Table II is that the elastic stiffness ($C_{22}$) along the zigzag (y) direction is also substantially smaller (around 2~3 times) than that of $MoS_2$ and GaSe. This is attributed to the intrinsic electronic properties of group IV monochalcogenides, whose covalence bonds are weaker than those of hexagonal TMDCs and group III monochalcogenides. This is also reflected in the longer bond lengths (2.50 ~2.89 Å) of our studied structures, compared with those of GaSe (2.47 Å) and MoS2 (1.84 Å) [14,16].

Recently, puckered 2D structures, such as few-layer black phosphorus (phosphorene), have attracted a significant amount of research attention. Due to their novel structure, unexpected mechanical properties have been shown to exist. In particular, phosphorene exhibits an unusually negative Poisson ratio[41]. Here we have calculated the Poisson ratio $\nu\perp$ obtained directly from relaxed ion coordinates by evaluating the change of layer thickness in response to in-plane hydrostatic strain $\Delta h/h = -\nu_{\perp}(\varepsilon_{11} + \varepsilon_{22})$. The Poisson ratio $\nu\perp$ is investigated by averaging the results of the armchair direction and zigzag direction for very small stress (-0.8% ~0.8%). Interestingly, our calculated value is positive and similar to those of TMDCs and group III monochalcogenides. This differs from the results of phosphorene, in which the Poisson ratio is evaluated by the value only from the armchair direction within a much larger stress range (-5% to 5%).[41]

*Piezoelectric coefficients:* Next, we calculate the linear piezoelectric coefficients of the group IV MX monolayers by evaluating the change of unit-cell polarization after imposing uniaxial strain on the system. This approach is based on the modern theory of polarization [24,25] implemented in VASP. The linear piezoelectric coefficients $e_{ijk}$ and $d_{ijk}$ are third-rank tensors as they relate polarization vector $P_i$, to strain $\varepsilon_{jk}$ and stress $\sigma_{jk}$ respectively, which are second-rank tensors:

$$e_{ijk} = \frac{\partial P_i}{\partial \varepsilon_{jk}} \qquad (3)$$

$$d_{ijk} = \frac{\partial P_i}{\partial \sigma_{jk}} \qquad (4)$$

Because of the existence of a mirror symmetry along the zigzag (y) direction in MX structures, the independent piezoelectric coefficients are {$e_{111}$, $e_{122}$, $e_{212}$=$e_{221}$} and {$d_{111}$, $d_{122}$, $d_{212}$ = $d_{221}$}. The indices 1 and 2 correspond to the *x* and *y* directions, respectively. The reason that $e_{212}$ = $e_{221}$ and $d_{212}$ = $d_{221}$ is because strain tensor is usually defined to be symmetric, namely $\varepsilon_{jk} = \varepsilon_{kj}$. The piezoelectric coefficients $e_{212}$ and $d_{212}$ describe the response of polarization to shear strain $\varepsilon_{12}$ and may be of less practical interest. Thus we will particularly focus on {$e_{111}$, $e_{122}$} and {$d_{111}$, $d_{122}$}, as well as the relationship between the $e_{ijk}$ and $d_{ijk}$.

By definition, the tensors are related by

$$e_{ijk} = \frac{\partial P_i}{\partial \varepsilon_{jk}} = \frac{\partial P_i}{\partial \sigma_{mn}} \frac{\partial \sigma_{mn}}{\partial \varepsilon_{jk}} = d_{imn} C_{mnjk} \tag{5}$$

where $C_{mnjk}$ are elastic constants. Einstein summation is implied for repeated indices. In 2D structures, an index can be either 1 or 2. Therefore,

$$e_{111} = d_{111} C_{1111} + d_{122} C_{2211} \tag{6}$$

$$e_{122} = d_{111} C_{1122} + d_{122} C_{2222} \tag{7}$$

Using the Voigt notation, we simplify it as $e_{11} = e_{111}$, $e_{12} = e_{122}$, $d_{11} = d_{111}$, $d_{12} = d_{122}$, $C_{11} = C_{1111}$, $C_{12} = C_{1122} = C_{2211}$. Then we can rewrite Eq. 6 and Eq.7 as

$$e_{11} = d_{11} C_{11} + d_{12} C_{12} \tag{8}$$

$$e_{12} = d_{11} C_{12} + d_{12} C_{22} \tag{9}$$

Furthermore, we can calculate $d_{11}$ and $d_{12}$ by $e_{11}$ and $e_{12}$ as

$$d_{11} = \frac{e_{11} C_{22} - e_{12} C_{12}}{C_{11} C_{22} - C_{12}^2} \tag{10}$$

$$d_{12} = \frac{e_{12} C_{11} - e_{11} C_{12}}{C_{11} C_{22} - C_{12}^2} \tag{11}$$

We have directly calculated the polarization of the MX monolayers by applying uniaxial strain $\varepsilon_{11}$ and $\varepsilon_{22}$ to the orthorhombic unit cell along the x direction and y direction, respectively. The change of polarization along y direction is zero because the mirror symmetry still remains under uniaxial strain for the $C_{2v}$ point group symmetric MXs.

The values of $e_{11}$ and $e_{12}$ are evaluated by a linear fit of 2D unit-cell polarization change along x direction ($\Delta P_1$) with respect to $\varepsilon_{11}$ and $\varepsilon_{22}$. In Figs. 2(a) and 2(b), we use $\varepsilon_{11}$ and $\varepsilon_{22}$ ranging from -0.005 to 0.005 in steps of 0.001 in the champed-ion case and -0.01 to 0.01 in steps of 0.002 in the relax-ion case. The dense steps of 0.001 are required for monolayer SnSe and SnS because their linear polarization changes occurring in the strain region are very small, less than ±0.004, as shown in Figs. 2(c) and 2(d). The fitting curves for these two materials are also confined with the linear region since the piezoelectric effect is a linear response according to strain. The relaxed-ion (or clamped-ion) $d_{11}$ and $d_{12}$ coefficients are finally calculated by the corresponding $e_{11}$, $e_{12}$ coefficients and elastic stiffness coefficients $C_{11}$, $C_{22}$ and $C_{12}$ based on Eq. 10 and Eq. 11.

We have summarized the calculated $e_{11}$, $e_{12}$, $d_{11}$, and $d_{12}$ coefficients in table III. The most useful piezoelectric coefficients (relaxed-ion $d_{11}$ and $d_{12}$), which reflect how much polarization charge can be generated with a fixed force and thus decide the mechanic-electrical energy converting ratio, are about 75 to 250 pm/V. Compared with those in frequently used bulk piezoelectric materials, such as α-quartz, wurtzite AlN, and ZnO, [7,14,42-44] and newly emerging 2D piezoelectric materials, such as $MoS_2$ and GaSe,[14,16,18] these values are about one to two orders of magnitude larger.

Finally, we find that the relaxed-ion $d_{11}$ and $d_{12}$ coefficients in the MX monolayers obey a periodic trend, as shown in Fig. 3. GeS possesses the smallest piezoelectric effect ($d_{11}$= 75.43 pm/V and $d_{12}$ = -50.42 pm/V), and moving upward in group 14 (crystallogens) and 16 (chalcogenide) enhances the magnitude of the effect until SnSe, which has the largest coefficient ($d_{11}$= 250.58 pm/V, $d_{12}$ = -80.31 pm/V), is reached. Interestingly, this trend is similar to that discovered in hexagonal TMDCs. [14]

These group IV monochalcogenide have highly desirable properties useful for a broad range of applications. In addition to this newly predicted giant piezoelectric effect, this family of single-atomic layers are flexible and transparent and can withstand enormous strain. On the other hand, for realistic devices, many other factors, in addition to the piezoelectric coefficients, will need to be considered. For instance, substrate effects and carrier mobilities are important for deciding the converting ratio in energy capture devices and the mechanical fatigue of these flexible materials has not been tested yet. These are beyond the scope of this Letter, but further research is expected in the near future.

*Conclusion:* By proven reliable tools, we have computationally demonstrated, for the first time, that monolayer group IV monochalcogenides MX, specifically GeS, GeSe, SnS, and SnSe, are intrinsically, giantly piezoelectric. The piezoelectric coefficients of this class materials are surprisingly one to two orders of magnitude larger than other frequently used piezoelectric materials. Encouraged by experimental achievements of monolayer samples, we expect that the huge piezoelectric properties of these materials to provide new platforms for electronic and piezotronic devices, and enable previously inaccessible avenues for sensing and control at the nanoscale.

AUTHOR INFORMATION

Corresponding Author

* lyang@physics.wustl.edu

ACKNOWLEDGEMENT

We acknowledge fruitful discussions with Luqing Wang, Vy Tran and Anders Carlsson. R. Fei. and L.Y. is supported by the National Science Foundation (NSF) Grant No. DMR-1207141 and NSF CAREER Grant No. DMR-1455346. JL and WBL acknowledge support by NSF DMR-1410636 and DMR-1120901. The computational resources have been provided by the Stampede of Teragrid at the Texas Advanced Computing Center (TACC) through XSEDE.

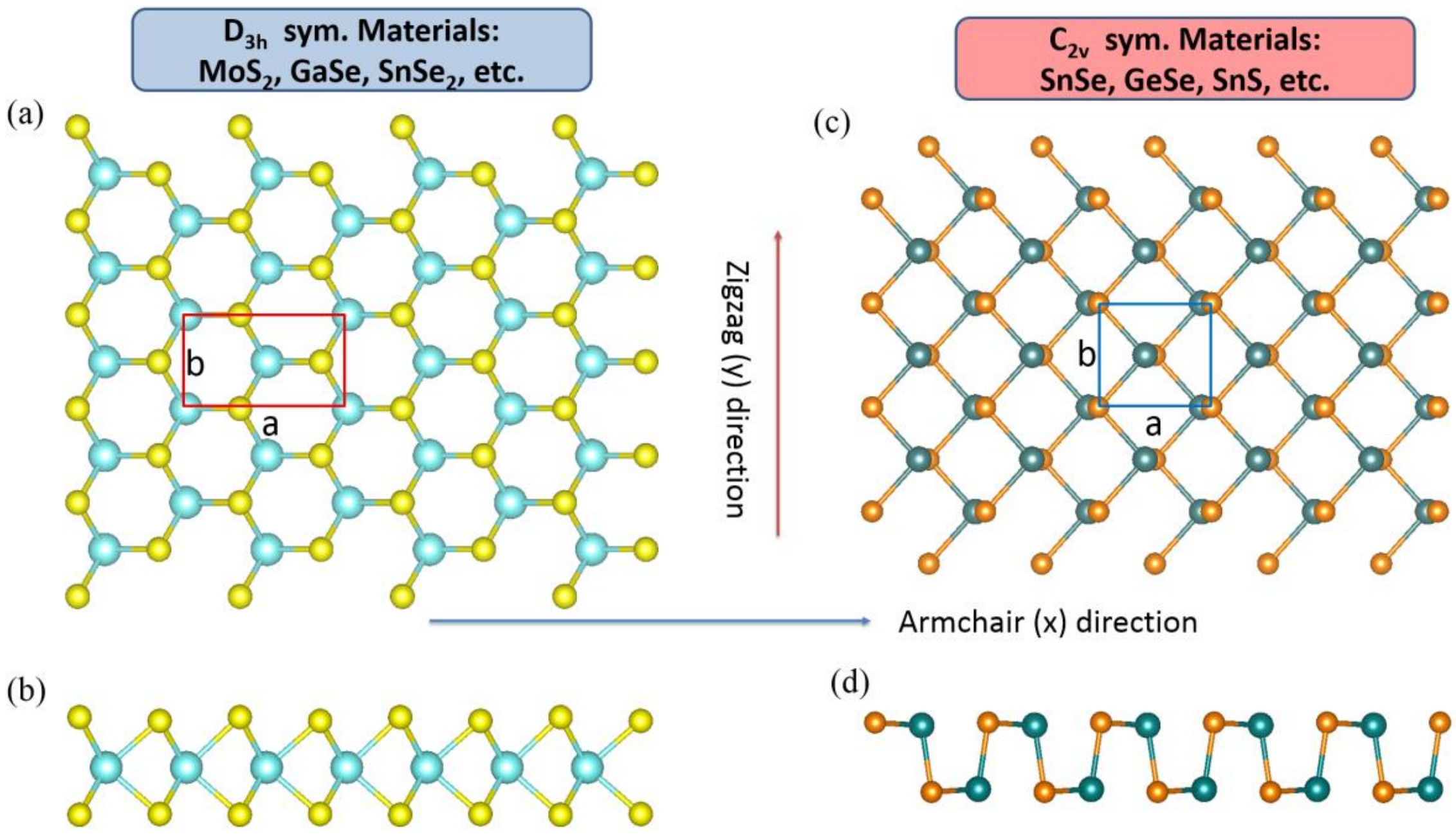

Figure 1 The ball-stick atomic structure of $D_{3h}$ hexagonal and $C_{2v}$ orthorhombic monolayers. (a) and (b) are the side top and side views of the hexagonal monolayer. (c) and (d) are the top and side views of the orthorhombic monolayer. The armchair direction and zigzag direction are defined as the x and y direction, respectively.

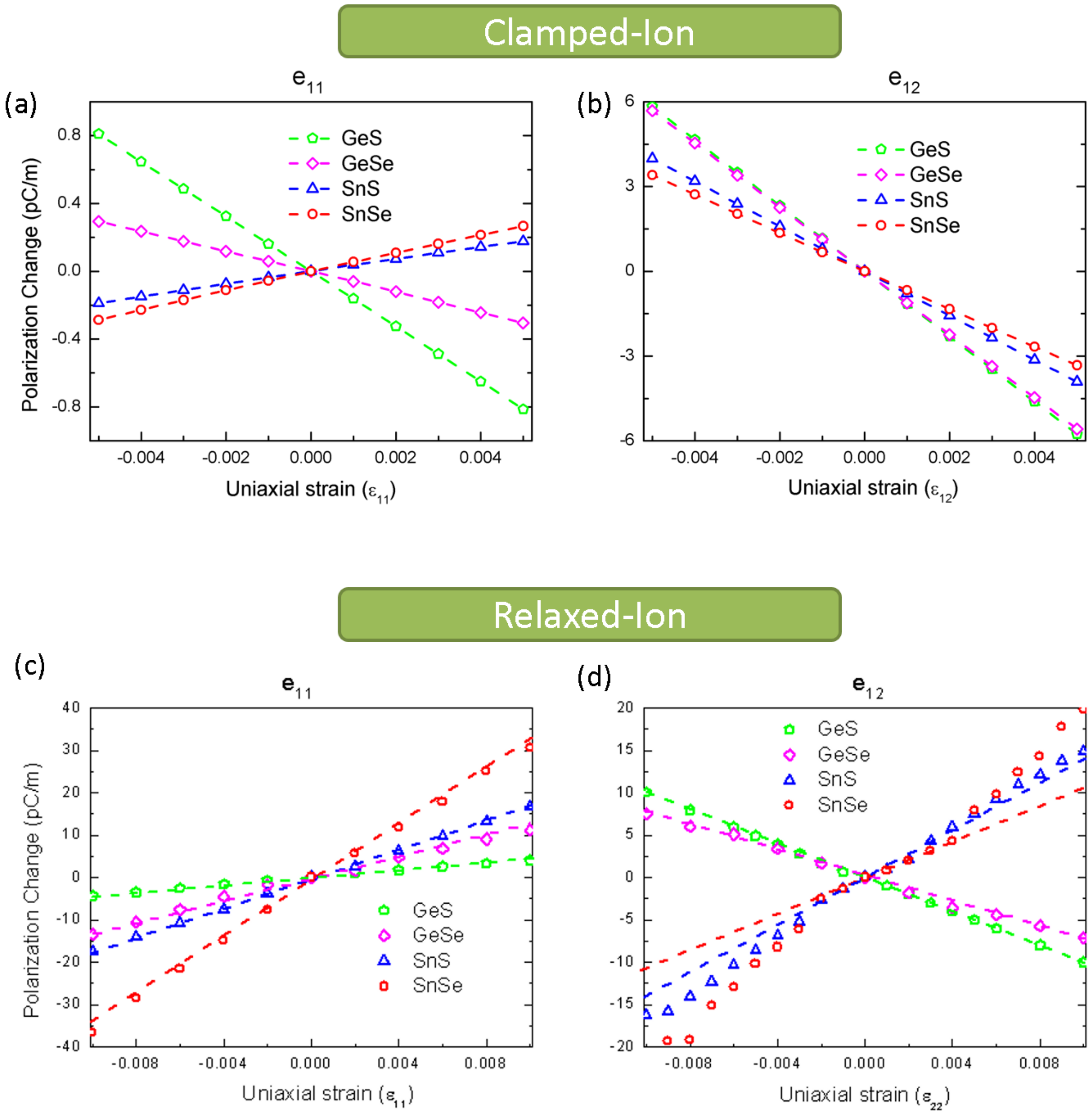

Figure 2 Change of unit-cell polarization per area of the MX monolayers along the x direction after applying uniaxial strain $\boldsymbol{\varepsilon}_{11}$ (a, c) and $\boldsymbol{\varepsilon}_{22}$ (b, d). Ionic positions within the unit cells were relaxed after imposing strain to the unit cell in the relaxed-ion case (c, d). The piezoelectric coefficients $\boldsymbol{e}_{11}$ and $\boldsymbol{e}_{12}$ correspond to the slope of lines obtained through linear fitting of polarization change with respect to $\boldsymbol{\varepsilon}_{11}$ and $\boldsymbol{\varepsilon}_{22}$.

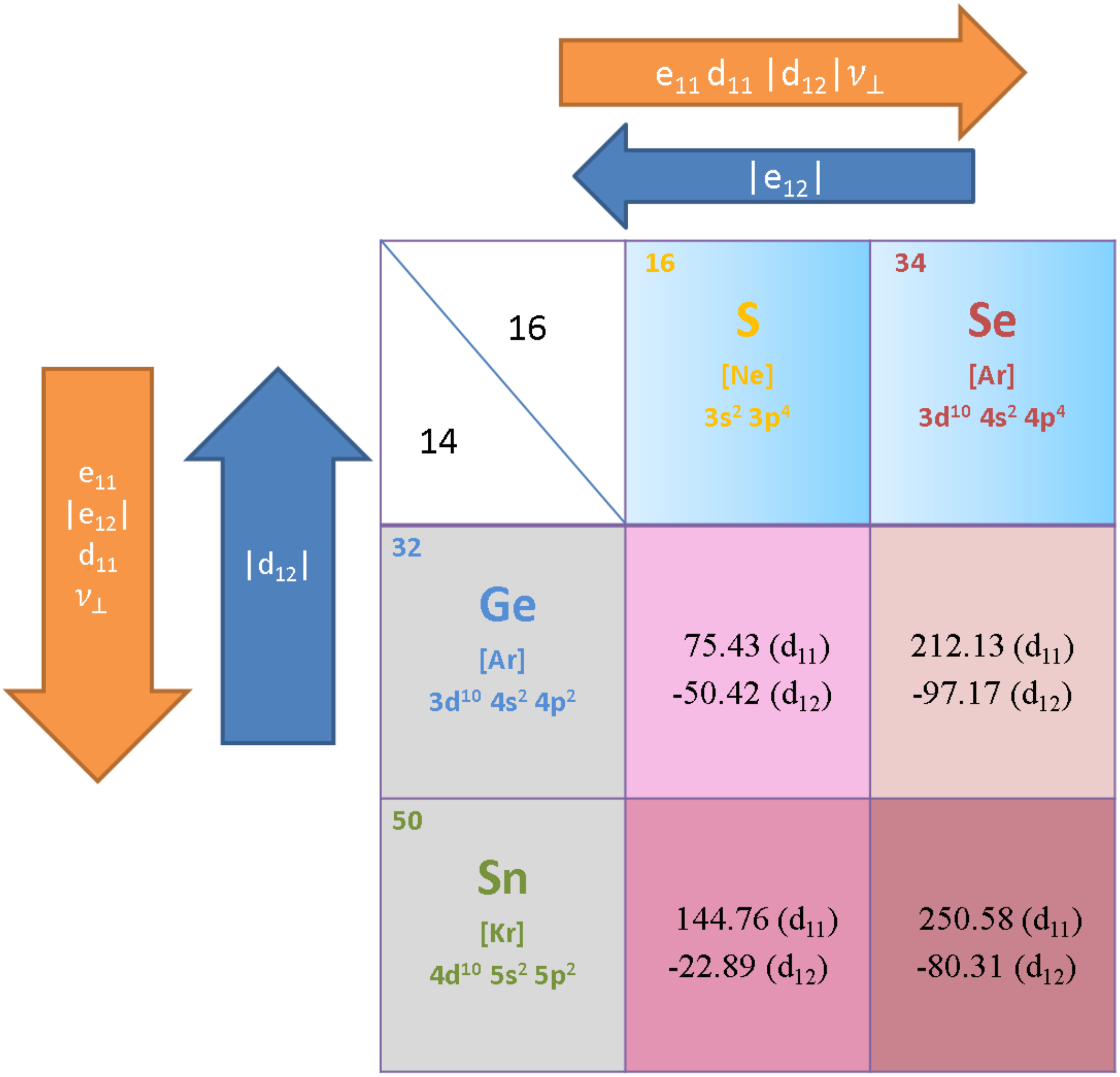

Figure 3 Trends of relaxed-ion structural, elastic, and piezoelectric properties of group IV MX monochalcogenides. The most practical interested relaxed-ion $d_{11}$ and $d_{12}$ coefficient values are listed as an example.

Table 1. Experimental and DFT-PBE calculated structural parameters and bandgap for bulk and monolayer MX. The values of monolayer lattice constants *a* and *b*, direct and indirect bandgap are listed.

| Material | monolayer DFT calculation | | | | bulk experiment or DFT calculation | | | |
|---|---|---|---|---|---|---|---|---|
| | a (Å) | b ( Å ) | Indirect gap (eV) | direct gap (eV) | a ( Å ) | b ( Å ) | Indirect gap (eV) | direct gap (eV) |
| **GeS** | 4.48 | 3.62 | 1.23 | 1.36 | 4.30 [28] exp. | 3.64 [28] exp. | 1.58 [29] exp. | 1.61 [29] exp. |
| **GeSe** | 4.27 | 3.93 | 1.04 | 1.10 | 4.38 [30] exp. | 3.82 [30] exp. | 1.16 [31] exp. | 1.53 [31] exp. |
| **SnS** | 4.26 | 4.03 | 1.37 | 1.51 | 4.33 [32] exp. | 3.99 [32] exp. | 1.07 [33] theory | 1.3 [34] exp. |
| **SnSe** | 4.35 | 4.24 | 0.77 | 0.92 | 4.44 [35] exp. | 4.14 [35] exp. | 0.86 [35] exp. | 1.30 [36] theory |

Table 2 DFT-PBE calculated in-plane elastic stiffness $C_{11}$, $C_{22}$ and $C_{12}$ of monolayer Group IV Monochalcogenides. The Poisson ratio $\nu_{\perp}$ is calculated for the relaxed ion case. The data of a typical TCMD monolayer material, $MoS_2$, and a typical group III monochalcogenide, GaSe, are listed for reference as well.

| Material | Clamp-ion | | | Relax-ion | | | |
|---|---|---|---|---|---|---|---|
| | $C_{11}$ (N/m) | $C_{22}$ (N/m) | $C_{12}$ (N/m) | $C_{11}$ (N/m) | $C_{22}$ (N/m) | $C_{12}$ (N/m) | $\nu_{\perp}$ |
| **GeS** | 48.90 | 58.19 | 32.92 | 20.87 | 53.40 | 22.22 | 0.32 |
| **GeSe** | 43.76 | 56.16 | 31.18 | 13.81 | 46.62 | 17.49 | 0.35 |
| **SnS** | 45.79 | 52.49 | 33.46 | 14.91 | 35.97 | 15.22 | 0.36 |
| **SnSe** | 43.96 | 47.60 | 30.66 | 19.88 | 44.49 | 18.57 | 0.42 |
| **$MoS_2$ [14]** | 153 | 153 | 48 | 130 | 130 | 32 | 0.34 |
| **GaSe [16]** | 108 | 108 | 32 | 83 | 83 | 18 | 0.39 |

Table 3 Calculated clamped-ion and relaxed-ion piezoelectric coefficients, $e_{11}$, $e_{12}$ , $d_{11}$, and $d_{12}$.

| **Material** | **Clamp-ion** | | | | **Relax-ion** | | | |
|---|---|---|---|---|---|---|---|---|
| | $e_{11}$ $10^{-10} C/m$ | $e_{12}$ $10^{-10} C/m$ | $d_{11}$ (pm/V) | $d_{12}$ (pm/V) | $e_{11}$ $10^{-10} C/m$ | $e_{12}$ $10^{-10} C/m$ | $d_{11}$ (pm/V) | $d_{12}$ (pm/V) |
| **GeS** | -1.62 | -11.6 | 16.39 | -29.21 | 4.6 | -10.1 | 75.43 | -50.42 |
| **GeSe** | -0.62 | -11.0 | 20.75 | -31.11 | 12.3 | -8.2 | 212.13 | -97.17 |
| **SnS** | 0.36 | -7.9 | 22.07 | -29.12 | 18.1 | 13.8 | 144.76 | -22.89 |
| **SnSe** | 0.65 | -6.68 | 20.46 | -27.21 | 34.9 | 10.8 | 250.58 | -80.31 |
| | | | | | | | | |
| **bulk α-quartz** | | | | | | | 2.3 [43] exp. | |
| **bulk AlN (wurtzite)** | | | | | | | 5.1 (d33)[44] exp. | |
| **ZnO** | | | | | 0.89 ($e_{33}$)[7] theory | -0.51 ($e_{31}$) [7] theory | 9.93 (d33)[42] exp. | |
| **MoS2** | | | | | 3.64[14] theory 2.9 [18] exp. | | 3.73[14] theory | |
| **GaSe** | 5.22 [16] theory | | 9.67 [16] theory | | 1.47 [16] theory | | 2.3 [16] Theory | |

Table of Contents

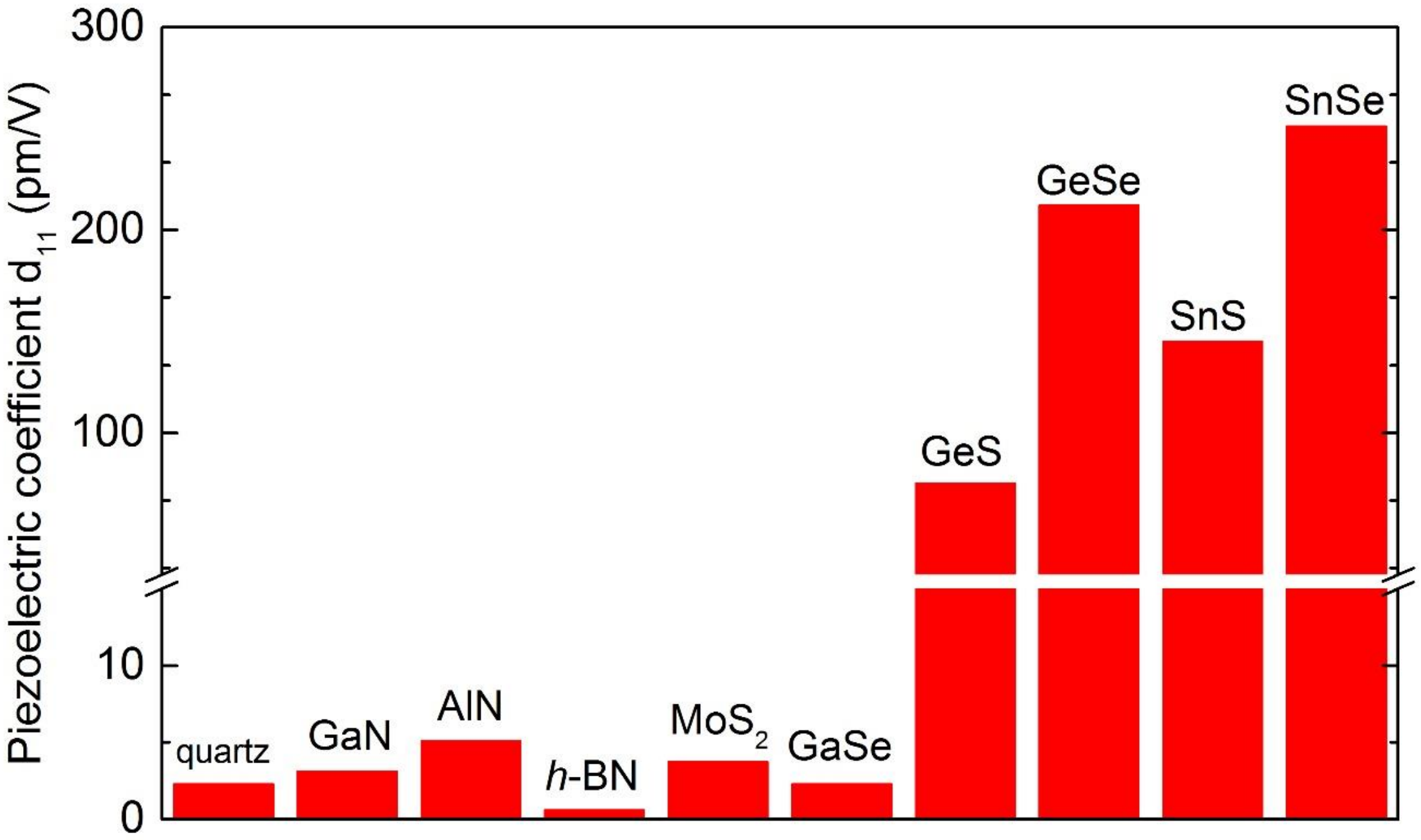